\documentclass[journal]{IEEEtran}

\usepackage{xcolor,soul,framed} 
\usepackage[justification=centering]{caption}

\colorlet{shadecolor}{yellow}
\usepackage[pdftex]{graphicx}
\graphicspath{{../pdf/}{../jpeg/}}
\DeclareGraphicsExtensions{.pdf,.jpeg,.png}

\usepackage{units}
\usepackage[cmex10]{amsmath}
\usepackage{array}
\usepackage{mdwmath}
\usepackage{mdwtab}
\usepackage{eqparbox}
\usepackage{url}
\usepackage{mathtools}
\usepackage{geometry}
\usepackage{xcolor}
\usepackage{cite}
\usepackage{multirow}
\usepackage{multicol}
\usepackage{booktabs}
\usepackage{subfig}
\usepackage[T1]{fontenc}
\usepackage{lipsum}
\usepackage[utf8]{inputenc}
\usepackage{array}
\usepackage{lettrine}
\usepackage{authblk}
\usepackage{graphicx}
\usepackage{stfloats}
\usepackage{amsmath}
\usepackage{epsfig}
\usepackage{amssymb}
\usepackage{comment}
\usepackage[utf8]{inputenc}
\usepackage[mathletters]{ucs}
\usepackage{tablefootnote}
\usepackage{algcompatible}
\usepackage{algorithm}
\usepackage{algpseudocode}
\usepackage{setspace}
\usepackage{bbm}
\usepackage[english]{babel}
\addto\captionsenglish{\renewcommand{\figurename}{Fig.}}
\usepackage{amsthm}
\usepackage{balance}

\usepackage[skip=2pt,font=footnotesize]{caption}

\begin{document}
   \title{Multi-Functional RIS for a Multi-Functional System: Integrating Sensing, Communication, and Wireless Power Transfer} \vspace{-0.3cm}
    	\newgeometry {top=25.4mm,left=19.1mm, right= 19.1mm,bottom =19.1mm}%
\author{Ahmed Magbool, Vaibhav Kumar, Ahmad Bazzi, Mark F. Flanagan, and Marwa Chafii
\thanks{This publication has emanated from research conducted with the financial
support of Science Foundation Ireland under Grant Number 13/RC/2077\_P2.}
\thanks{Ahmed Magbool and Mark F. Flanagan are with the School of Electrical and Electronic Engineering, University College Dublin, D04 V1W8 Dublin, Ireland (e-mail: ahmed.magbool@ucdconnect.ie; mark.flanagan@ieee.org).}
\thanks{Vaibhav Kumar, Ahmad Bazzi, and Marwa Chafii are with the Engineering Division, New York University Abu Dhabi, UAE. Ahmad Bazzi and Marwa Chafii are also with NYU WIRELESS, NYU Tandon School of Engineering, Brooklyn, NY, USA (e-mail: vaibhav.kumar@ieee.org; ahmad.bazzi@nyu.edu; marwa.chafii@nyu.edu).}}

\maketitle

\begin{abstract}
Communication networks are evolving from solely emphasizing communication to facilitating multiple functionalities. In this regard, integrated sensing, communication, and powering (ISCAP) provides an efficient way of enabling data transmission, radar sensing, and wireless power transfer simultaneously. Such a multi-functional network requires a multi-functional architectural solution. Toward this end, sensor-aided zero-energy reconfigurable intelligent surfaces (SAZE-RISs) offer an energy-efficient solution for ISCAP by meeting the requirements of the end users as well as supplying power for the RIS. This paper explores the use of SAZE-RIS within the ISCAP framework. First, we present the general system architecture, operational protocols, and main application scenarios for employing SAZE-RIS in ISCAP. Next, we discuss methods for managing the conflicting requirements of communication, sensing, and powering within ISCAP and the role of SAZE-RIS in this process. We then provide a detailed case study complete with simulation results, offering valuable insights into the design choices and tradeoffs that come into play when adopting this technology. Furthermore, we discuss the related challenges and open research avenues, highlighting areas that require further exploration to fully realize the potential of SAZE-RIS within this ISCAP framework.
\end{abstract}

\IEEEpeerreviewmaketitle
\section{Introduction} \label{sec:intro}
Communication networks are evolving to support multiple functions beyond that of communication~\cite{2023_Chafii}. Sixth-generation (6G) networks, for example, are expected to offer not only high data rates (10-100 Gbps on average and up to 1 Tbps peak) but also precise indoor and outdoor positioning accuracy (1 cm and 50 cm on average, respectively) and a 20-year battery life for low-energy internet-of-things (IoT) devices~\cite{2024_Yilong}.

At the architectural level, reconfigurable intelligent surfaces (RISs) are emerging as a promising technology for 6G networks. These surfaces, made up of nearly-passive meta-elements, can dynamically adjust the phase and sometimes the amplitude of the incident signals~\cite{2019_Basar}. Traditionally used in communication to create adaptive non-line-of-sight (NLoS) links, RISs have shown potential in focusing signal energy toward specific directions. This capability has led researchers to explore their use in various applications, including wireless sensing~\cite{2022_Zhang} and wireless power transfer (WPT)~\cite{2022_Tran}. 

Further studies have also explored the use of RISs to support multiple functions simultaneously, particularly in integrated sensing and communication (ISAC), a paradigm where key resources such as hardware, spectrum, and signal processing are shared between communication and radar sensing, leading to more efficient resource utilization~\cite{2022_Liu}. RISs have proven effective in improving both communication and radar performance by manipulating signal propagation, increasing integration gains, and reducing tradeoffs between communication and sensing, as demonstrated in~\cite{2024_Magbool} and the references therein. Additionally, sensor-aided RISs, which incorporate active sensors for radar sensing, have been proposed in multiple works (e.g.~\cite{2022_Yu}) to strengthen echo signals by eliminating the need for signals to travel back and forth between the base station (BS) and the target via the RIS.

Integrated sensing, communication, and powering (ISCAP) is an advanced multifunctional approach that combines WPT with ISAC~\cite{2024_Yilong}. The utilization of RIS to support the multiple functionalities in  ISCAP has been explored in a number of works, such as~\cite{2024_Li} and~\cite{2024_Chen}. A key application of ISCAP is the wireless transfer of power to operate RISs, leading to the development of zero-energy RISs (ZeRISs). These RISs capture part of the incident signal to power themselves while reflecting the remaining signal to serve the receivers~\cite{2023_Tyrovolas}. This approach is viable because RIS elements require minimal power, making WPT an effective solution.

Building on the previous concepts, this paper discusses employing sensor-aided zero-energy RIS (SAZE-RIS) within the ISCAP framework, while also examining the associated use cases, opportunities and challenges, and charting future research directions. The discussion commences by presenting the general system architecture, the operational protocols, and prominent application scenarios for employing SAZE-RIS for ISCAP in Section~\ref{sec:sys_model}. Following this, Section~\ref{sec:opt} discusses the ways of managing the conflicting requirements for SAZE-RIS-assisted ISCAP systems. A case study and simulation results are presented in Section~\ref{sec:case_study}, providing valuable insights. Subsequently, Section~\ref{sec:challanges} explores associated challenges and outlines open research avenues. Finally, concluding remarks are provided in Section~\ref{sec:conc}.

\section{System Architecture and Application Scenarios} \label{sec:sys_model}

In this section, we outline the system architecture of SAZE-RIS for ISCAP and discuss the operational protocols and main application scenarios.

\subsection{System Architecture}

\begin{figure*}
         \centering
         \includegraphics[width=2\columnwidth]{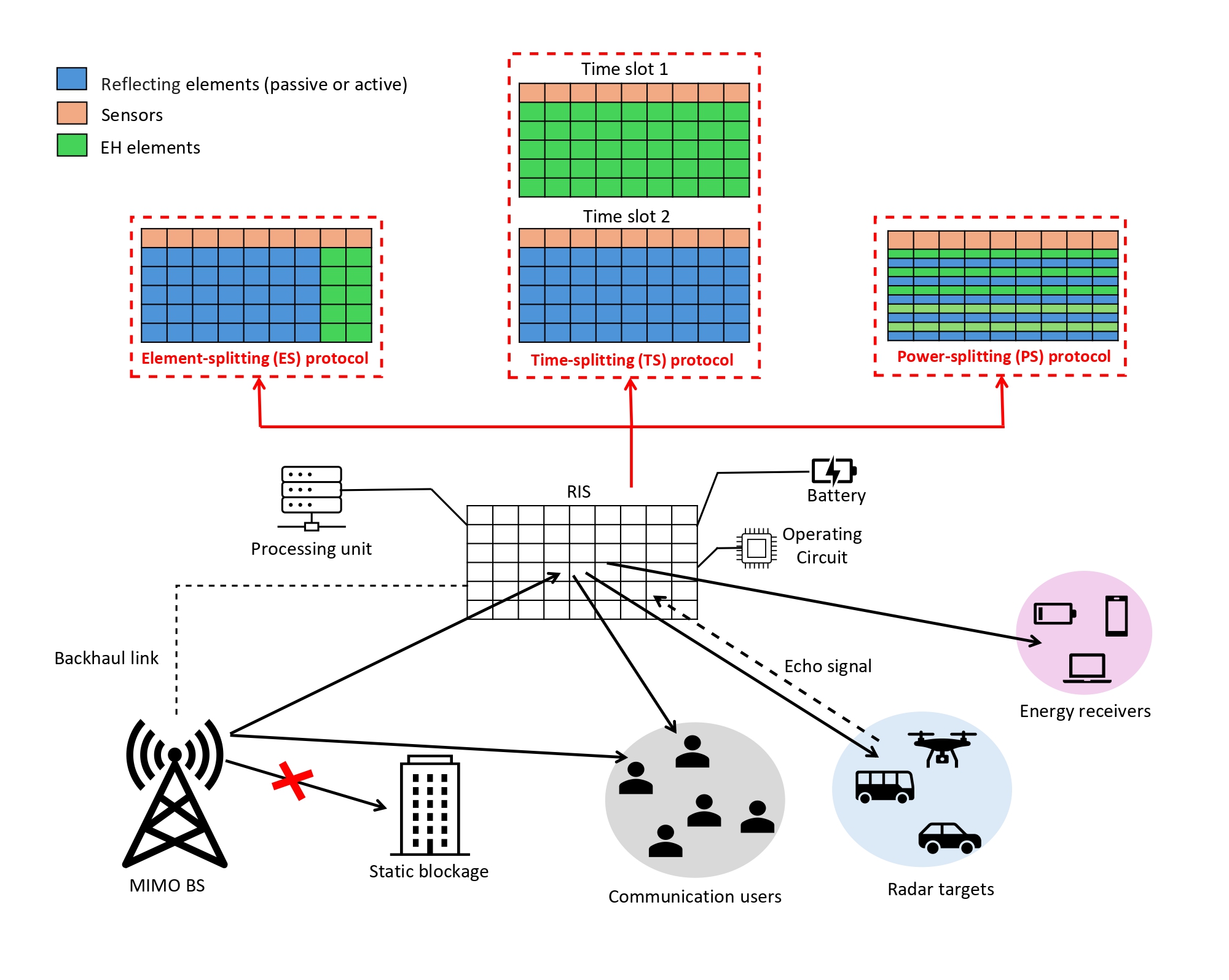}
        \caption{System architecture and transmission protocols for SAZE-RIS-assisted ISCAP.}
        \label{fig:sys_model}
\end{figure*}

Fig.~\ref{fig:sys_model} illustrates the general system architecture and transmission protocols for SAZE-RIS-assisted ISCAP. In this setup, a multiple-input multiple-output (MIMO) BS transmits an ISCAP signal\footnote{More details about the ISCAP transmit waveform will be presented in Section~\ref{sec:wfd}).} with the aim of simultaneously transmitting data to a cluster of users, sensing multiple targets, and supplying power to a number of energy receivers with the assistance of a SAZE-RIS.

The SAZE-RIS consists of multiple elements, each capable of fulfilling one or more of three primary functions. 
\begin{itemize}
    \item \textbf{Reflecting elements} to redirect the signal towards the receivers, either with or without amplitude amplification.
    \item \textbf{Radar sensors} to sense the reflected echoes from radar targets. These echo signals are then forwarded to a processing unit for further analysis. Due to the randomness in communication data that might reduce the accuracy of echo processing, the presence of a backhaul link between the BS and the SAZE-RIS is essential. This link could be established either wirelessly or through wired means.
    \item \textbf{Energy harvesting (EH) elements} to partially absorb the energy of the incident signal(s) and store it in a battery to power the SAZE-RIS.
\end{itemize}

A SAZE-RIS differs from a conventional RIS in several ways. First, a SAZE-RIS is designed to operate with minimal or no external power by harvesting energy from the incident signals, unlike a conventional RIS that typically relies on a dedicated power supplies. Furthermore, a SAZE-RIS incorporates sensors for sensing, leading to greater operational autonomy and better quality echo signals. This integration of sensing and energy harvesting also enables SAZE-RIS to prioritize communication, sensing, EH or WPT depending on the current conditions. The specific design of SAZE-RIS for ISCAP makes it ideal for scenarios where power sources are limited or where autonomous operation is beneficial, offering a sustainable and efficient solution.

\subsection{Transmission Protocol}
To facilitate the operation of SAZE-RIS, one of three protocols can be adopted. These protocols are detailed below and summarized in Table~\ref{tab:RIS_protocols}.

\begin{table*}
\footnotesize
\centering
\caption{Operational protocols for the multi-functional RIS. The example system used in the table below consists of a 128-element SAZE-RIS. On average, each transmission involves 4 communication users, 2 radar targets, and 2 energy receivers. Additionally, each EH element is capable of powering 3 RIS elements per time unit.}
\begin{tabular} {| m{2cm} || m{4.7cm} | m{4.7cm} | m{4.7cm} |} 
 \hline  Protocol & Element-splitting & Time-splitting & Power-splitting   \\  [0.5ex] 
 \hline
  \hline
 Description & The SAZE-RIS elements are divided into three distinct groups: one group operates as reflecting elements, another as EH elements, and the remaining elements function as sensors.  &  The SAZE-RIS elements are divided into two distinct groups: one group operates as sensors, while the others function as reflecting elements for specific time slots and as  EH elements for the other time slots. & The SAZE-RIS elements are divided into two distinct groups: one group operates as sensors, while the others store some part of the signal towards the EH and reflect the other part for the communication and energy receivers.  \\
  \hline
 Protocol division parameter & The number of elements in each group is designed based on long-term statistical modeling of the system. & The time slots allocated for EH and reflecting are determined by solving an optimization problem to optimize the performance of the system based on short-term system modeling. & The energy of the signals being absorbed and reflected by the RIS can be optimized based on the RIS energy requirement and short-term system setup. \\
 \hline
 Hardware design & Requires a single-function circuit per element.  & Requires a switch and two circuits per switching element. & Requires a circuit capable of storing energy and reflecting signals simultaneously. \\
  \hline
 Division of the example system described in the caption of the table & 64 SAZE-RIS elements are utilized as reflecting elements, benefiting communication users, radar targets, and energy receivers. Additionally, 32 SAZE-RIS elements are allocated as EH elements to power the other 96 elements, while the remaining 32 elements are designated as sensors, benefiting radar targets. & 32 elements are designated as sensors while the other 96 elements function as EH elements for 1/4 of the time and reflecting elements for 3/4 of the time. This allows the SAZE-RIS to be self-powered. & 32 elements are utilized as sensors, while the remaining 96 are used for signal reflection and SAZE-RIS powering. In each of these 96 elements, 25\% of the power is absorbed to power the SAZE-RIS, and 75\% is reflected to the receiver. \\

 \hline
\end{tabular}
\label{tab:RIS_protocols}
\end{table*}

\textbf{1. Element-splitting (ES) protocol:} In the ES protocol, SAZE-RIS elements are divided into three distinct groups: reflecting elements, EH elements, and sensor elements, which operate simultaneously. This segmentation simplifies circuit design by assigning each element a specific task. It also eliminates the need to fine-tune the division parameters for each transmission. However, this approach offers limited flexibility in adapting to changing system conditions. Optimal division can still be achieved by prioritizing certain functionalities based on statistical models that account for the behavior and locations of communication users, radar targets, and energy receivers.

\textbf{2. Time-splitting (TS) protocol:} The TS protocol alternates the function of a set of SAZE-RIS elements between reflecting and EH roles in different time slots, while a separate group remains as dedicated sensor elements. This approach provides greater flexibility compared to the ES protocol, allowing for optimization of instantaneous performance metrics for communication and SAZE-RIS powering. However, it requires a more complex hardware design, with switches needed to connect elements to one of two circuits.

\textbf{3. Power-splitting (PS) protocol:} In the PS protocol, SAZE-RIS elements split the energy of the incoming signals from the BS, with part of the signal absorbed for EH and the rest reflected to the receivers. This setup allows the energy harvesting and reflecting elements to work simultaneously, while the sensor elements operate in an ES mode. The division of tasks is controlled by a PS parameter, which can be adjusted to optimize specific performance metrics, offering flexibility similar to the TS approach. However, the PS protocol introduces challenges, including increased hardware complexity due to the need for splitting, reflecting, and EH circuits for each element, and added complexity in solving resource allocation problems when incorporating the PS parameter.

\subsection{Application Scenarios}
\begin{figure}
    \centering
    \includegraphics[width=1\linewidth]{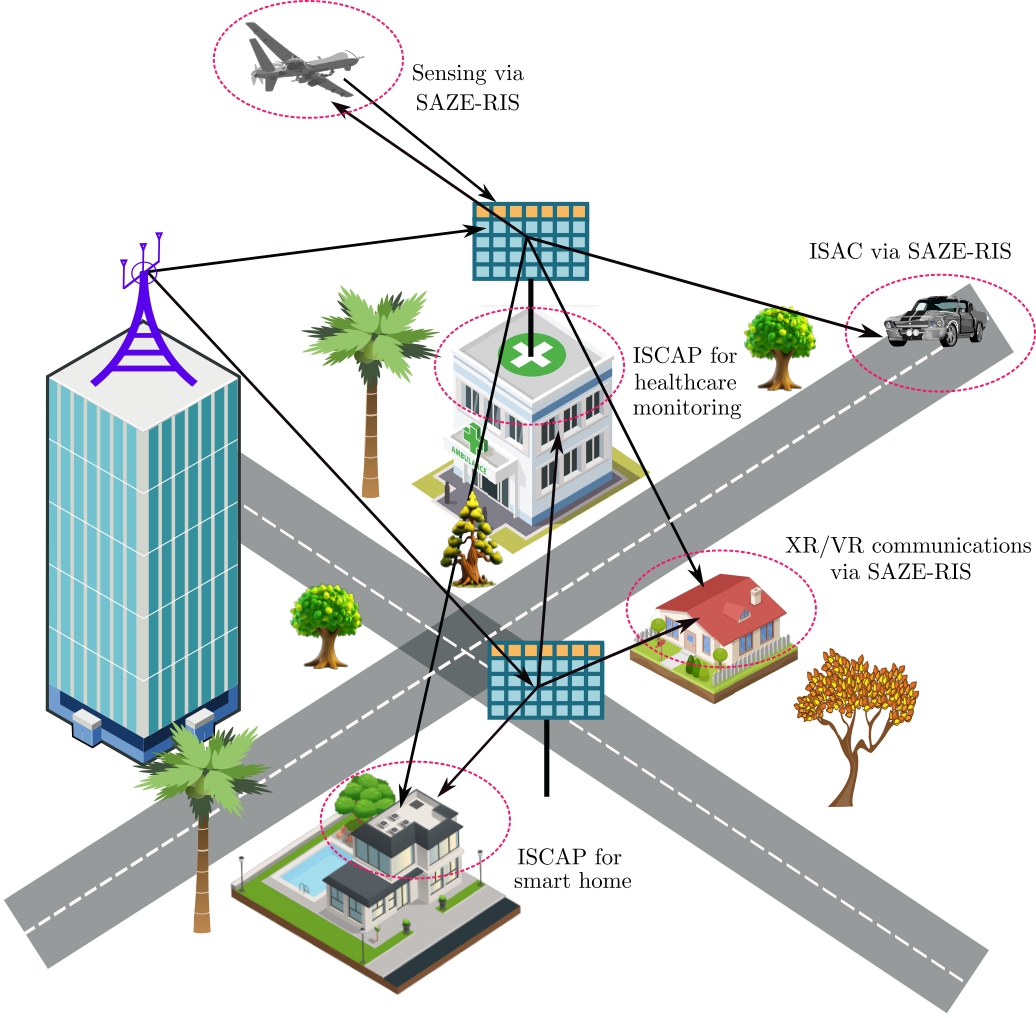}
    \caption{Application scenarios for SAZE-RIS-enabled ISCAP systems.}
    \label{fig:applications}
\end{figure}

\textbf{1) Sensing:} As depicted in Fig.~\ref{fig:applications}, a SAZE-RIS can improve sensing accuracy in several ways. Without a line-of-sight (LoS) link between the BS and the target, SAZE-RIS's sensor elements can process the echo signal directly, enhancing the sensing performance. For example, \cite{2024_Qiaoyan} reported a 70 dB improvement in the Cramér-Rao bound (CRB) for direction-of-arrival (DoA) estimation with a sensor-aided RIS compared to a conventional RIS. Even with a LoS link, SAZE-RIS can act as an additional sensing node for greater accuracy.

\textbf{2) ISAC:} SAZE-RIS can boost ISAC performance and reduce the communication-sensing tradeoff. Notably, it enhances channel reconstruction through location sensing, as proposed in~\cite{2023_Hu}. By leveraging the accurate sensing capabilities of SAZE-RIS, user locations are determined for precise channel estimation, eliminating the need for pilot-based overhead and freeing up resources for data transmission, thus improving system efficiency and throughput.

\textbf{3) Extended reality (XR) and virtual reality (VR):} In XR and VR environments, real-time data transmission is essential for an immersive experience. SAZE-RIS can support high data rates, ensuring users receive seamless, high-resolution visuals and low-latency audio. Its precise sensing capabilities can also enhance the user experience by accurately tracking user movements and interactions, resulting in more natural and responsive interactions within the virtual space. Additionally, SAZE-RIS improves energy efficiency, helping to extend the battery life of VR devices and controllers, which is crucial for prolonged immersive sessions without frequent recharging interruptions.

\textbf{4) Smart homes:} With ISCAP, smart homes can process sensor data in real time, enabling advanced automation, security monitoring, energy management, and personalized services. Distributed SAZE-RIS units, deployed at various locations, can serve as localization anchors. Additionally, these SAZE-RIS units can simultaneously power IoT devices and themselves, enhancing energy efficiency within the smart home environment. 

\textbf{5) Healthcare monitoring:} Healthcare monitoring is a key application for ISAC, as identified by the 3rd Generation Partnership Project (3GPP). SAZE-RISs, when placed in healthcare facilities or homes, can ensure seamless data transmission from wearable sensors and medical devices, such as heart rate monitors, glucose sensors, and blood pressure cuffs. They also function as localization anchors, providing accurate tracking of patients' positions and movements for precise monitoring and quick emergency responses. Additionally, the self-powering capability of SAZE-RISs enhances system sustainability by reducing the need for frequent battery replacements and maintenance.

\section{Managing the Conflicting Requirements of SAZE-RIS-Assisted ISCAP} \label{sec:opt}
Multi-functional systems often encounter conflicting requirements, complicating optimization and performance measurement. In SAZE-RIS-assisted systems, two key tradeoffs emerge. First, energy tradeoff between the energy absorbed for harvesting versus the energy reflected for communication or sensing. Second, functionality tradeoff between optimizing the communication, sensing, and WPT at the receivers.

This section discusses strategies for managing these conflicts in SAZE-RIS-assisted ISCAP systems, covering waveform design, optimization methods, and tradeoffs, and shows how SAZE-RIS can help addressing these challenges.

\subsection{Waveform Design for ISCAP}  \label{sec:wfd}
In ISCAP system, the transmitted signal must simultaneously support communication, sensing, and energy harvesting functionalities, which complicates the waveform design process. There are two ways to design ISCAP waveforms.

\textbf{1. Orthogonal design:} To manage transmission for communication, sensing, and energy harvesting in ISCAP, the available spectrum or time resources can be divided using techniques like frequency-division multiplexing (FDM) or time-division multiplexing (TDM). However, this division can reduce the resources available to each function, potentially affecting performance by lowering spectral efficiency for communication, reducing resolution for sensing, and decreasing received power for energy harvesting.

\textbf{2. Joint design:} To address reduced performance in orthogonal designs, joint waveforms must be carefully designed to simultaneously support communication, sensing, and EH. This can be achieved by combining signals through superposition or developing new waveforms via optimization. However, integrating multiple functionalities into one signal can cause cross-function interference, such as radar pulses interfering with communication data, which can degrade both functions. Additionally, the waveform needs to be optimized to maximize energy capture while still meeting communication and sensing requirements.

\subsection{System Optimization}
Optimizing ISCAP systems is complex due to the different performance metrics for sensing, communication, and WPT. Consider as an example an ISCAP system with a set of resources we can control, such as transmit and receive beamformers and RIS phase shifts. System constraints, including power budgets and constant modulus constraints at the RIS, should be taken into account. Additionally, for self-powered RISs with TS protocol (or PS protocol), another set of constraints exists, governing the time (or energy) allocated for RIS EH and for signal reflection. When aiming to maximize metrics for communication, sensing, and WPT simultaneously, a multi-objective optimization problem arises. This problem is challenging because the three objectives often have conflicting trends, making it difficult to find a single optimal solution for all of them.

While there are several approaches to optimization problems with multiple objective functions, we present three common methods.

\begin{itemize}
    \item \textbf{Keeping one objective function and constraining the others:} The first approach to optimizing ISCAP systems involves optimizing only one objective function while setting thresholds for the others to ensure their minimum performance levels are met. This method faces two main issues. First, stringent thresholds can limit feasible solutions, which in turn compromises the objective function. Second, the additional performance constraints, which are often non-convex and nonlinear, can increase the complexity and difficulty of finding an optimal solution.
    \item \textbf{Lexicographic approach:} To address the challenge of choosing suitable threshold values that do not overly restrict the feasible region, the lexicographic approach can be used, as exemplified in~\cite{2023_Magbool}. This method involves ranking the three objective functions by importance, such as prioritizing communication first, then sensing, and finally WPT. The lexicographic approach then solves three optimization problems sequentially, focusing on each objective function in order of priority.
    
    The first optimization problem focuses on optimizing the communication performance metric while adhering to the original constraints, ignoring sensing and WPT. The second optimization problem maximizes the sensing performance metric with the original constraints and an additional constraint on the allowable reduction in the communication performance from its optimal value obtained in the first stage. The third optimization problem optimizes the WPT metric with the same constraint set as the second problem and an additional constraint on the allowable reduction in the sensing metric from its optimal value. While this approach helps achieve a more balanced solution for all objectives, it is computationally demanding as it requires solving three separate optimization problems.
    \item \textbf{Creating a new objective function:} The third option involves crafting a new objective function, which is a composite function of the three performance metrics. A commonly utilized objective function is the weighted sum due to its simplicity. The optimization problem then aims to optimize the new objective function subject to the original constraint set. Although the objective function becomes more intricate, the feasible region in this scenario is broader than in the previous two cases since no extra constraints are included. Nonetheless, determining the appropriate composite objective function can be quite challenging, particularly when the various performance metrics are non-homogeneous, such as combining communication spectral efficiency, radar detection probability, and harvested power by the energy receivers.
\end{itemize}

After choosing a method to address the multi-objective optimization problem, it is essential to use suitable tools to attain the optimal solution while balancing performance and complexity. The gradient projection and Riemannian optimization methods are common methods for optimizing the RIS phases due to their efficiency in handling the bounded modulus constraint. Other frequently used optimization techniques to address such problems include semi-definite relaxation (SDR), successive convex approximation (SCA), and penalty-based methods.

Machine learning (ML) and reinforcement learning (RL) are increasingly used to tackle multi-objective optimization problems. ML techniques, such as supervised and unsupervised learning, can model system behavior based on historical data, enhancing the accuracy and efficiency of optimization across multiple objectives. Neural networks, for example, can approximate Pareto-optimal solutions or identify patterns for better trade-offs in SAZE-RIS assisted ISCAP. RL excels in dynamic scenarios, using trial-and-error to develop strategies that balance multiple objectives. Techniques like Q-learning, Deep Q-Networks, and policy gradient methods enable real-time optimization, allowing the system to adapt to changing conditions while balancing multiple objectives.

\subsection{Tradeoffs}
\begin{figure}
         \centering
         \includegraphics[width=0.99\columnwidth]{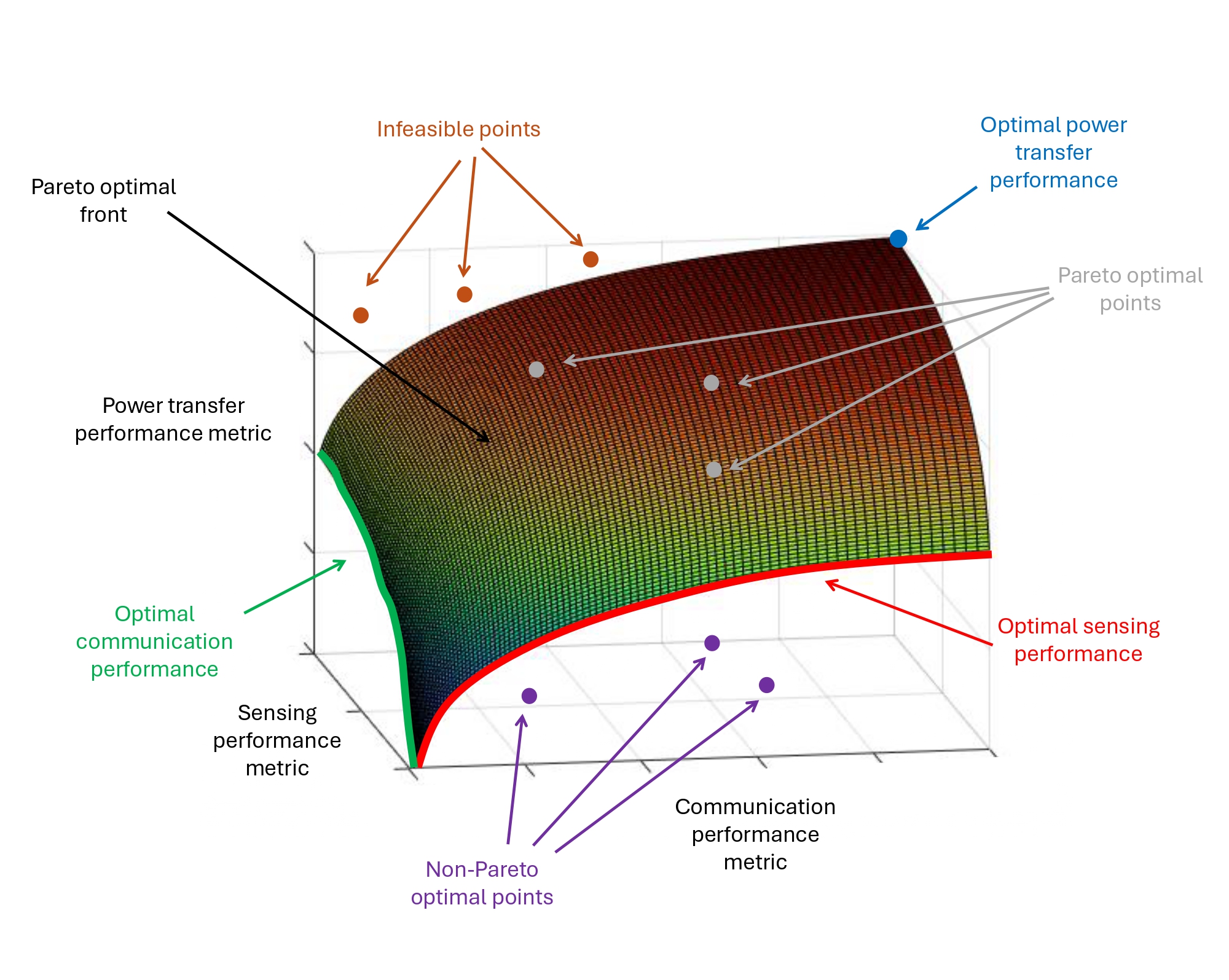}
        \caption{The Pareto optimal front of the integrated sensing, communication, and wireless power transfer problem.}
        \label{fig:Pareto_opt}
\end{figure}

Managing the conflicting requirements of different systems results in tradeoffs. For example, consider a communication performance metric (e.g., spectral efficiency), a sensing metric (e.g., detection probability), and a WPT performance metric (e.g., harvested power). The goal is to allocate resources to achieve an operational point that is close the maximum values of all these metrics. Fig.~\ref{fig:Pareto_opt} offers valuable insights into this process, which we explain below.

\begin{itemize}
    \item \textbf{Pareto optimality:} In the previous subsection, we introduced three methods to tackle the ISCAP multi-objective optimization problem, each involving specific thresholds or control parameters. Adjusting these parameters produces the \textit{Pareto optimal front}, as illustrated in Fig.~\ref{fig:Pareto_opt}. Points on this front signify optimal performance for one function (e.g., sensing) given the performance metrics of other functions (e.g., communication and WPT). It is important to note that different Pareto optimal fronts can emerge depending on the method used to tackle the multi-objective optimization problem, some of which are sub-optimal (i.e., can be outperformed by other Pareto optimal fronts at every point).
    \item \textbf{Function optimality:} The optimal performance of one function (e.g., sensing) when the other functions are disregarded is represented as a point or a curve on the Pareto optimal front. Ideally, the Pareto optimal front will have points that are simultaneously close to all the three individual optimal function points.
    \item \textbf{Infeasible points:} Infeasible points lie above the Pareto optimal front and cannot be reached using the adopted method. For example, achieving both high communication spectral efficiency and high radar detection probability simultaneously is impossible for certain situations because improving one degrades the other.
    \item \textbf{Non-Pareto feasible points:} These points are situated below the Pareto optimal front and, while feasible, are undesirable as operational points. However, they may be located on another sub-optimal Pareto front.
\end{itemize}

\subsection{Role of SAZE-RIS in Managing the Conflicting Requirements of ISCAP}
In addition to its capability of providing energy-efficient solutions for ISCAP, SAZE-RIS has high flexibility which can manage the conflicting requirements of ISCAP. In this subsection, we provide a number of examples of this.

\textbf{1. Propagation control for functionality prioritization:} 
SAZE-RIS acts as an additional resource to optimize and balance the conflicting demands of communication, sensing, and powering requirements. Using advanced algorithms and real-time adjustments, SAZE-RIS can dynamically prioritize these functions based on current needs. For example, it can enhance communication during peak traffic by optimizing signal paths for higher throughput, then shift its focus to energy harvesting and sensing during quieter periods to ensure efficient resource use.

\textbf{2. Improving the Pareto-optimal front:} Different problem formulations and solution techniques can produce various Pareto-optimal fronts. An ideal Pareto-optimal front shows a correlation between the three functions, meaning the optimal communication performance is close to that of the sensing and WPT ones. SAZE-RIS can enhance this by improving the coupling between different systems, thereby enhancing Pareto-optimality. This concept was explored for ISAC in~\cite{2023_Meng}, where the problem was formulated to maximize the coupling between communication and sensing sub-spaces. This approach can similarly be extended to include the WPT sub-space.

\textbf{3. Sensing-assisted communication and powering by SAZE-RIS:} Both communication and powering can benefit from location information. Accurate channel state information is crucial for effective data transmission in communication, while locating energy receivers beforehand is essential for WPT. The embedded sensors in the SAZE-RIS have proven effective for channel estimation, removing the need for dedicated pilots, as demonstrated in~\cite{2023_Hu}. This approach can also be utilized for localizing energy receivers.

\section{Case Study}    \label{sec:case_study}
In this section, we present a case study on the optimal beamforming design for a SAZE-RIS-aided ISCAP system from a joint design perspective. Specifically, we aim to minimize the total transmit power from the BS while considering the quality-of-service (QoS) requirements for communication, sesning, WPT and EH as optimization constraints.

\subsection{Setup}
We consider a system where an $N_{\mathrm t}$-antenna BS communicates with two single-antenna information receivers (IRs), transfers power wirelessly to two single-antenna energy receivers (ERs), and simultaneously senses a target. The SAZE-RIS is equipped with $N_{\mathrm s}$ sensors to sense the target, and $N_{\mathrm{r}}$ elements for energy harvesting and/or signal reflection. In the case of the PS protocol, $\rho$ denotes the fraction of the received power at the SAZE-RIS used for signal reflection, and in the case of the TS protocol, $\rho$ represents the fraction of the time slot used for signal reflection at the SAZE-RIS. Moreover, in the case of ES, $\rho$ denotes the fraction of the total number of SAZE-RIS elements used for signal reflection, i.e., $\rho N_{\mathrm r} \triangleq N_{\mathrm{reflect}}$, while the rest of the elements are dedicated for EH, i.e., $(1 - \rho)N_{\mathrm r} = N_{\mathrm r} - N_{\mathrm{reflect}} \triangleq N_{\mathrm{harvest}}$. $P_{c}$ denotes the constant circuit-power consumption at the RIS (for all three operating protocols), and $P_{\mathrm e}$ denotes the power consumed by each of the reflecting elements at the RIS. Therefore, the total power required to be harvested at the SAZE-RIS in the PS protocol is given by $P_{\mathrm{PS}} = P_{\mathrm c} + \rho N_{\mathrm r} P_{\mathrm e}$, that in the TS protocol is given by $P_{\mathrm{TS}} = P_{\mathrm c} + \rho N_{\mathrm r} P_{\mathrm e}$, and that in the ES protocol is given by $P_{\mathrm{ES}} = P_{\mathrm c} + N_{\mathrm{reflect}} P_{\mathrm e}$. We formulate a joint beamforming optimization problem to minimize the total transmit power from the BS while guaranteeing a sensing rate of $0.2$~bps/Hz at the sensors, a communication rate of  $1$~bps/Hz at each of the IRs, and a total harvested power of $0.5$~mW at the ERs. It is assumed that the power harvesting efficiency at the RIS/ERs is 0.8, the mean radar cross-section (RCS) of the target is 0.5~$\text{m}^2$, the number of sensors is $N_{\mathrm s} = 10$, $P_{\mathrm e} = 2 \ \mu$W, and $P_{\mathrm c} = 50$~mW. 

A stationary solution to the formulated non-convex optimization problem is obtained using an alternating optimization (AO)-based approach, where in each iteration, the optimal receive beamforming vector at the sensors is obtained in closed-form using the generalized Rayleigh quotient and eigenvalue decomposition, while we simultaneously update the transmit beamforming vectors and the RIS reflection beamforming vector using the SCA method followed by solution of a second-order convex program (SOCP). 
%

\begin{figure}[t]
    \centering
    \includegraphics[width=0.8\linewidth]{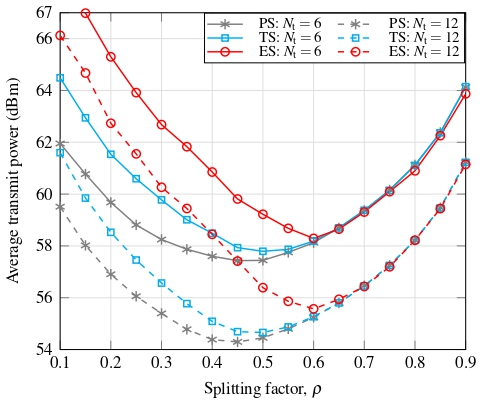}
    \caption{Impact of the splitting factor ($\rho$) on the average transmit power requirement with $N_{\mathrm r} = 100$.}
    \label{fig:vary_rho}
\end{figure}

\subsection{Results and Analysis} 
\begin{figure}[t]
    \centering
    \includegraphics[width=0.8\linewidth]{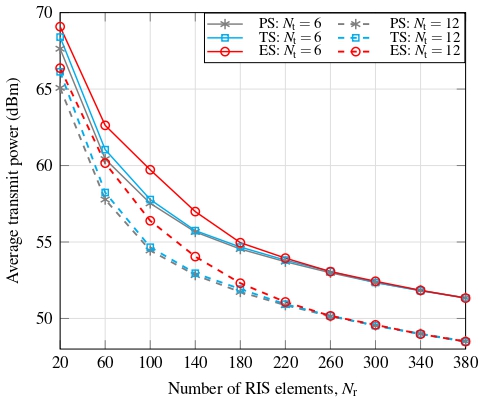}
    \caption{Impact of the number of RIS elements ($N_{\mathrm r}$) on the average transmit power requirement with $\rho = 0.5$.}
    \label{fig:vary_nR}
\end{figure}

Fig.~\ref{fig:vary_rho} illustrates the impact of the splitting factor $\rho$ on the average transmit power requirement for the system, where an increase in the number of transmit antennas $N_{\mathrm t}$ leads to a decrease in the required average transmit power due to the active beamforming gain at the BS. Notably, there exists a non-trivial optimal value of $\rho$ for each operating protocol, which needs to be configured for optimal system performance. For the PS protocol, when $\rho$ is small, a small fraction of the received signal power at the RIS is used for signal reflection, while a large fraction is harvested for RIS operation. As $\rho$ increases, the average transmit power requirement decreases due to a larger fraction of power used for signal reflection, but beyond a certain $\rho$ value, a very small fraction of signal power remains available for harvesting at the RIS. The system performance in this case is restricted by the power harvesting constraint at the RIS, resulting in a larger transmit power requirement. The system's behavior for the TS and ES protocols can be explained in a similar manner. In Fig.~\ref{fig:vary_nR}, we show the impact of the number of RIS elements on the system's performance for $\rho = 0.5$. It can be noted from the figure that with an increase in the number of RIS elements, the beamforming gain offered by the RIS increases, resulting in a smaller required transmit power. 

\vspace{0.25cm}
\section{Challenges and Open Research Directions} \label{sec:challanges}

Despite its potential, SAZE-RIS-assisted ISCAP requires further research for full integration into 6G systems. This section highlights key research challenges and future directions for this technology.

\subsection{Hardware}
Designing circuits for SAZE-RIS across all protocols requires a pragmatic approach. For instance, in the PS protocol, circuits must handle both EH and signal reflection. This can be done using hot-carrier diodes for RF-to-DC conversion and supercapacitors for power storage. Another major hardware challenge is integrating EH elements that support wideband harvesting, as traditional circuits operate within narrow resonance bands, limiting harvesting. Thus, hardware design should focus on EH circuits that work across wide resonance bands for optimal EH.

\subsection{Security \& Privacy Concerns}
In addition to communication users, radar targets, and energy receivers, unauthorized access to sensing, communication, and powering is a concern for ISCAP systems enabled by SAZE-RIS technology. To address this, additional degrees-of-freedom for security are required in the design process. For instance, artificial noise can be used as part of a resource set to reduce eavesdropping potential. Changes to the Pareto optimal frontier depicted in Fig.~\ref{fig:Pareto_opt} can be expected for security reasons, and these points can serve as tighter benchmarks for secure RIS-assisted ISCAP systems. For example, if an eavesdropper is a target of interest for sensing purposes, the system can impose constraints on communication and harvested power while maintaining a threshold sensing performance.

More serious inherent security concerns could arise when a sophisticated eavesdropper exploits periods when the SAZE-RIS is operating on low energy levels. In particular, an attacker can make full use of the energy unpredictability by timing attacks to coincide with periods when the device is operating on low energy. An open challenge here is to properly manage SAZE-RIS ISCAP resources to account for potential malicious attacks. Besides security, privacy is another major concern for SAZE-RIS, where ISCAP information can be leaked to third-party users in order to learn different patterns, thus facilitating malicious activity. 

\subsection{Near-field Operation}
Not only are future wireless systems anticipated to operate at smaller and smaller wavelengths, but also applications entailing large numbers of RIS elements can also contribute in enlarging the near-field region. For this, near-field ISCAP has to account for the wavefront curvature in geometric near-ﬁeld conditions in order to properly fulfill its near-field ISCAP tasks. Nevertheless, we can expect a larger number of Pareto fronts as near-field channels tend to convey more degrees of freedom due to the higher rank of near-field channels. In turn, future near-field ISCAP systems should exploit near-field channels in order to make full use of the multiplexing capabilities offered by these channels.

\subsection{Machine Learning}
ML can effectively contribute in the design of ISCAP systems, given their multifunctional nature. For instance, in the TS protocol mode, greedy ML methods can determine the EH versus reflecting mode. Additionally, if large datasets of channel state information are available across network nodes, including the MIMO BS and SAZE-RIS, supervised learning approaches can be beneficial for estimating the SAZE-RIS phase shifts and predicting the optimal splitting factor $\rho$. Generative ML, capable of generating new data based on observed patterns, can aid in generating RIS phases (and amplitudes for active RIS), alongside receive and transmit beamformers.

\subsection{Backhauling and Self-Interference}
While wired backhauling is popular, especially for indoor small cells, wireless backhauling offers low-cost maintenance and scalability advantages. Standardized by 3GPP Rel-16 and Rel-17, the European Commission wireless backhauling has garnered significant research interest for efficient solutions.

A key challenge with SAZE-RIS is managing limited backhaul capacity, often seen with low-cost solutions. When using such backhaul links, traffic to and from the SAZE-RIS must be restricted. The dynamic nature of sensing and communication data complicates the efficient representation and compression of ISCAP information. An alternative is to develop an advanced optimization approach to handle ISCAP tasks within the constraints of limited backhaul capacity.

SAZE-RIS faces self-interference as it simultaneously reflects data, senses targets, and powers receivers, which can degrade performance. To mitigate this, phase adjustments can be made to reduce the interference of reflected signals with the received echo signal.

\subsection{Sensing-Assisted Communication and powering}
While sensing may be required as a separate service, it is valuable to explore how sensing information can be utilized for communication and EH purposes. For instance, SAZE-RIS-assisted ISCAP can leverage sensing to enhance communication and EH performance, becoming more channel-aware due to radar processing. However, a significant challenge arises when performing ISCAP tasks under conditions where the central processing unit has access only to inaccurate sensing information, compounded by the dynamic nature of the environment and additional complications like moving obstacles and clutter.
As LoS channels are fundamentally ideal for sensing and powering, while NLoS channels are more suited for communications, proper accurate modeling for SAZE-RIS is crucial to capture ISCAP system performance.

\balance
\section{Conclusion}   \label{sec:conc}
This paper provided an overview of the use of SAZE-RIS technology as a framework for designing ISCAP systems. It commenced by outlining the fundamental system architecture, operational protocols, and key application scenarios where SAZE-RIS can be effectively employed within ISCAP systems. Subsequently, strategies for reconciling the conflicting demands of communication, sensing, and powering within ISCAP were presented, emphasizing the pivotal role of SAZE-RIS in this regard. A comprehensive case study was then discussed, augmented with simulation results, providing valuable insights into the implementation of this technology. Moreover, the associated challenges and promising research directions were presented to unlock the full potential for emerging applications in this space.



%





\ifCLASSOPTIONcaptionsoff
  \newpage
\fi





\bibliographystyle{IEEEtran}
\bibliography{IEEEabrv,Bibliography}
%


\end{document}